\documentclass[aps,prb,twocolumn,showpacs,10pt,floatfix,longbibliography]{revtex4-2}

\usepackage{newtxtext,newtxmath}
\usepackage{graphicx}
\usepackage{subfigure}
\usepackage{amsmath}
\usepackage{amsfonts}
\usepackage{mathrsfs}
\usepackage{bbm}
\usepackage{subfigure}
\usepackage{xcolor}
\usepackage[colorlinks=true]{hyperref}
\usepackage[T1]{fontenc}

\usepackage{siunitx}
\usepackage{bm}

\usepackage[version=4]{mhchem}

\begin{document}

\title{Tunneling valley Hall effect induced by coherent geometric phase}

\author{W.~Zeng}
\email[E-mail: ]{zeng@ujs.edu.cn}
\affiliation{Department of physics, Jiangsu University, Zhenjiang 212013, China}

\begin{abstract}
We propose a geometric phase-resolved tunneling valley Hall effect based on the coherent transmission through two combined electric barriers in $\alpha-\mathcal{T}_3$ lattices. It is shown that the backreflected electrons at the barrier interface may acquire a valley-dependent geometric phase. The coherence of this geometric phase leads to the valley-dependent skew tunneling, which is responsible for the transverse valley current with zero net charge. We further demonstrate that this charge-neutral transverse valley Hall current can be electrically controlled by the gate voltages applied across the two combined barrier regions and is absent when the two barriers are of equal height. Our work opens a new approach to generating the valley Hall effect, suggesting potential applications for valleytronic devices.

\end{abstract}
\maketitle

\section{Introduction}\label{intro}
In optics, when a polarized light beam returns to its initial state via two intermediate polarizations, it acquires a phase shift known as the Pancharatnam phase \cite{pancharatnam1956generalized,doi:10.1080/09500348714551321,PhysRevLett.126.183902,PhysRevLett.108.190401}. This phase is determined by the geodesic triangle whose vertices are the three polarizations on the Poincar\'e sphere \cite{PhysRevA.109.023517,2005progress}. As an analogy to the polarized light, the electron with spin (pseudospin) can acquire an electron version of the Pancharatnam phase over the course of a cycle process, depending on the geometric path taken through the Bloch sphere, which is known as the geometric phase \cite{anandan1992geometric,RevModPhys.64.51,RevModPhys.66.899}. Specifically, when the wave vector completes one rotation around the $\mathbf{k}=0$ point, this geometric phase is equal to the number of rotations of the pseudospin, namely the pseudospin winding number or Berry phase \cite{PhysRevB.84.235432,PhysRevB.84.205440,PhysRevLett.120.087402}. Contrary to the usual Berry phase, the geometric phase can be acquired by the single electron tunneling process at the interface of the heterojunctions and electrically tuned by the junction control, leading to the  nontrivial charge and spin transport, such as the Veselago lens formed in a graphene nanoribbon \cite{PhysRevB.89.155412}, topological waveguides and topological transistor \cite{PhysRevB.87.165420}. However, the discussions on the valley-dependent geometric-phase devices are still insufficient in the literature.

\begin{figure}[tp]
\begin{center}
\includegraphics[clip = true, width =\columnwidth]{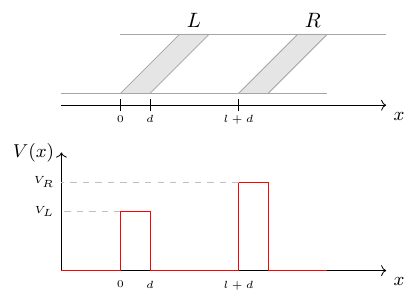}
\end{center}
\caption{Schematic illustration of the $\alpha-\mathcal{T}_3$ lattice based junction (top panel). The longitudinal direction of the junction is along the $x$ axis. The left ($L$) and right ($R$) electrostatic potential barriers of width $d$ are indicated by the gray regions. The width between $L$ and $R$ is $l$. The potential profile of the tunneling junction is shown in the bottom panel, where the heights of the $L$ and $R$ barriers are $V_L$ and $V_R$, respectively. }
\label{fig:junction}
\end{figure}

Valleytronics is an emerging device concept in condensed matter physics, offering novel ways to manipulate information beyond traditional charge and spin-based methods \cite{PhysRevLett.124.037701,PhysRevB.98.201407,PhysRevB.87.155415,PhysRevB.105.094510,PhysRevLett.132.156301}. In analogy with the spin Hall effect \cite{PhysRevLett.96.106802,PhysRevLett.83.1834,RevModPhys.87.1213}, the valley Hall effect is a hot topic attracting much attention in the literature \cite{PhysRevB.92.125146,PhysRevB.103.195309,PhysRevB.104.075421}. Similar to the skew scattering on impurities, the skew tunneling in heterojunctions is an alternative mechanism to generate the anomalous Hall effect in the ballistic regime, which is termed as the \textit{tunneling} Hall effect \cite{PhysRevLett.115.056602,PhysRevB.106.L180404}. Distinct from the intrinsic mechanism, which is directly linked to the band topology and where the Hall conductance is proportional to the integration over the Fermi sea of the Berry curvature of each occupied band \cite{PhysRevLett.99.236809,PhysRevLett.75.1348,RevModPhys.82.1959}, the tunneling Hall effect originates from the presence of asymmetric chiral contributions to the transmission probability. The tunneling spin Hall effect has been predicted in the heterojunctions with spin-orbit coupling barriers \cite{PhysRevLett.115.056602,PhysRevB.106.L180404,PhysRevB.100.060507}. Very recently works reported that the tunneling valley Hall effect can be generated by tilting the Dirac cones \cite{PhysRevLett.131.246301,PhysRevLett.132.096302}. All the aforementioned approaches to generating the tunneling spin (valley) Hall effect are Berry-curvature-free but require breaking the spatial symmetry along the transverse direction of the tunneling junctions.

In this paper, we propose a new method for the generation of the tunneling valley Hall effect based on the geometric phase coherent tunneling in $\alpha-\mathcal{T}_3$ lattices. The $\alpha-\mathcal{T}_3$ lattice is an extension of the graphene honeycomb lattice with an additional site centered at each hexagonal cell \cite{PhysRevB.99.205429,PhysRevB.99.205135,PhysRevB.106.094503,PhysRevB.92.245410}. The low-energy excitations in the $\alpha-\mathcal{T}_3$ lattice are the massless pseudospin-one Dirac fermions characterized by the valley-contrasting $\alpha$-dependent Berry phase, where the parameter $\alpha$ provides a continuous lattice transformation from the graphene-like lattice ($\alpha=0$) to the dice lattice ($\alpha=1$). Several methods have been proposed to realize the $\alpha-\mathcal{T}_3$ lattices in experiments, such as the $\ce{SrTiO3/SrIrO3/SrTiO3}$ trilayer heterostructure grown along the ($111$) \cite{PhysRevB.84.241103} direction and the $\ce{Hg_{1-x}Cd_{x}Te}$ at the critical doping \cite{PhysRevB.92.035118}. We show that the backreflected electrons in $\alpha-\mathcal{T}_3$ lattices may acquire a valley-dependent geometric phase through the single tunneling process. This acquired geometric phase is determined by the height of the barrier, which can be electrically controlled. The geometric phase coherent transport can be generated by combining two successive single barriers coherently \cite{datta1997electronic}, \textit{i.e.}, the Fabry-P\'erot model \cite{PhysRevB.83.155440,PhysRevB.75.045103,PhysRevLett.81.5370}, as shown in Fig.~\ref{fig:junction}. We show that the incident angle resolved net transmission probability of the Fabry-P\'erot interferometer is asymmetric for a given valley, leading to a nonzero transverse valley Hall current. However, the total transmission probability ($K$ valley $+$ $K^\prime$ valley) is symmetric due to the time-reversal symmetry. As a result, the transverse charge current is always zero. We further demonstrate that this charge-neutral transverse valley Hall current can be electrically controlled by the gate voltages applied across the left ($V_L$) and right ($V_R$) barrier regions, and disappears at $V_L=V_R$.

The remainder of the paper is organized as follows. The model Hamiltonian and the scattering approach are explained in detail in Sec.\ \ref{model}. The numerical results and discussions are presented in Sec.\ \ref{results} . Finally, we conclude in Sec.\ \ref{conclusions}.

\section{theoretical Model}\label{model}
The system under consideration is shown in Fig.\ \ref{fig:junction}, where the longitudinal direction of the $\alpha-\mathcal{T}_3$ lattice-based junction is along the $x$ axis. The electrons tunneling through the two combined electric barriers are described by the low-energy Hamiltonian \cite{PhysRevLett.112.026402,PhysRevB.92.245410,PhysRevB.92.155116}
\begin{align}
    \mathcal{H}=\begin{pmatrix}
        0&f_\tau(\mathbf{k})\cos\varphi&0\\
        f^*_\tau(\mathbf{k})\cos\varphi&0&f_\tau(\mathbf{k})\sin\varphi\\
        0&f^*_\tau(\mathbf{k})\sin\varphi&0
    \end{pmatrix}+V(x),\label{eq:hamiltonian}
\end{align}
where the lattice parameter is $\varphi=\tan^{-1}\alpha$, $\tau=\pm1$ labels the $K$ and $K^\prime$ valleys, respectively, $f_\tau(\mathbf{k})=\hbar v_F(\tau k_x-ik_y)$ with $v_F$ being the Fermi velocity and $\mathbf{k}=(k_{x},k_y)^T$ being the wave vector in the $x-y$ plane. The gate-voltage controlled electrostatic potential is given by
\begin{equation}
   V(x)=
    \left\{
        \begin{array}{ll}
         V_L, & 0<x<d, \\
          V_R, & l+d<x<l+2d, \\
        0, &  \text{others},
        \end{array}
    \right.
\end{equation}
where $V_L$ ($V_R$) is the gate potential applied on the left (right) barrier, $d$ is the barrier width and $l$ is the width between two barriers; see Fig.~\ref{fig:junction} (bottom panel). 

In the absence of the electrostatic potential, the spectrum consists of two linearly dispersing branches with a flat band cutting through the Dirac point, which is given by $E=s|\mathbf{k}|$ with $s=\{+1,-1,0\}$ denoting the conduction, valence and flat band, respectively. The Berry connection of each band can be obtained by $\mathbf{A}_{s,\tau}=i\langle\mathbf{k},s,\tau|\bm\nabla_{\mathbf{k}}|\mathbf{k},s,\tau\rangle$ with the corresponding eigenstate $|\mathbf{k},s,\tau\rangle=(\tau\cos\varphi e^{-i\tau\theta},s,(-1)^{s+1}\tau\sin\varphi e^{i\tau\theta})^T$, where $\tau=\pm1$ is the valley index and $\tan\theta=k_y/k_x$. The Berry phase is given by $\Phi_{s,\tau}=\oint_\mathcal{C}d\mathbf{k}\cdot\mathbf{A}_{s,\tau}$ with $\mathcal{C}$ being any closed path encircling the degeneracy point in the momentum space. Consequently, one finds that the Berry phase for the conical bands ($s=\pm1$) and the flat band ($s=0$) are $\Phi_\tau=\tau(1-\alpha^2)/(1+\alpha^2)\pi$ and $\Phi_{\tau,0}=-2\Phi_\tau$, respectively \cite{PhysRevLett.112.026402}. For $\alpha=0$, the Berry phase is $\Phi_\tau=\pi$ and the Hamiltonian describes the graphene system with an extra inert flat band. For $\alpha=1$, the Hamiltonian describes the massless pseudospin-one systems with a vanishing Berry phase. For $\alpha\neq0,1$, the valley-contrasting non-$\pi$ Berry phase appears.

The scattering states can be obtained by the secular equation
\begin{align}
\mathcal{H}\psi(x)e^{ik_yy}=E\psi(x)e^{ik_yy},\label{eq:sch}
\end{align}
where $k_y$ and $E$ are the conserved transverse wave vector and the incident energy, respectively. The wave function $\psi(x)$ is dependent on the electrostatic potential $V(x)$. In the barrier regions, where $V(x)=V_p$ with $p=\{L,R\}$ distinguishing the left and right barriers, respectively, the scattering basis states are given by
\begin{align}
    \mathbf{u}_{p}^> = 
    \begin{pmatrix}
    \tau e^{-i\tau\theta_p}\cos\varphi\\
    1\\
   \tau e^{i\tau\theta_p}\sin\varphi\end{pmatrix},\quad
        \mathbf{u}_{p}^< = 
    \begin{pmatrix}
    \tau e^{i\tau\theta_p}\cos\varphi\\
    -1\\
    \tau e^{-i\tau\theta_p}\sin\varphi\end{pmatrix}.\label{eq:states}
\end{align} 
Here the superscript `$>(<)$' denotes the right (left) propagating direction and the transmitted angle $\theta_{p}$ satisfies $\sin\theta_{p}=\hbar v_Fk_y/(E-V_{p})$. We note that the factors $e^{\pm ik_px}$ are omitted in Eq.\ (\ref{eq:states}) for simplicity, where $k_p=(E-V_p)\cos\theta_p/\hbar v_F$ is the longitudinal wave vector. In the pristine region with $V(x)=0$, the left (right) propagating basis states $\mathbf{u}_{0}^>$ ($\mathbf{u}_{0}^<$) can be obtained by the replacement $V_p\rightarrow0$ in Eq.\ (\ref{eq:states}), where the incident angle ($\theta_0$) and the longitudinal wave vector ($k_0$) satisfy $\sin\theta_0=\hbar v_Fk_y/E$ and $k_0=E\cos\theta_0/\hbar v_F$, respectively.

The probability current $\mathbf{j}$ can be obtained from the continuity equation $\partial_t|\psi|^2+\bm\nabla\cdot\mathbf{j}=0$ with the wave function $\psi=(u_1,u_2,u_3)^T$ satisfying Eq.\ (\ref{eq:sch}), which yields
\begin{align}
j_x=v_F\mathrm{Re}[u_2^*(u_1\cos\varphi+u_3\sin\varphi)].
\end{align}
Consequently, the conservation of $j_x$ at the barrier interface requires the continuity of $u_1\cos\varphi+u_3\sin\varphi$ and $u_2$ \cite{PhysRevB.95.235432,PhysRevB.105.165402,PhysRevB.103.165429}, which results in the transfer matrix for the $p$ barrier ($p=L,R$)
\begin{gather}
\mathcal{M}_p=\mathcal{O} e^{ik_pd\sigma_z}\mathcal{O}^{-1},\quad \mathcal{O}=(\bar{\mathbf{u}}_p^>,\bar{\mathbf{u}}_p^<).\label{eq:tM}
\end{gather}
Here $\sigma_z$ is the $z$-component of the spin-$1/2$ Pauli matrix. The new two-component basis state $\bar{\mathbf{u}}$ in Eq.\ (\ref{eq:tM}) is given by $\bar{\mathbf{u}}=(u_1\cos\varphi+u_3\sin\varphi,u_2)^T$ with $u_i$ ($i=1,2,3$) being the $i$-th component of $\mathbf{u}$ in Eq.\ (\ref{eq:states}). The transmission probability for the single barrier $p$ ($p=L,R$) is determined by the transfer matrix $\mathcal{M}_p$, which is given by
\begin{align}
t_p=\big[\bm{\Gamma}\bar{\mathbf{u}}_0^>(\bar{\mathbf{u}}_0^>)^\dagger\bm{\Gamma}^\dagger\big]_{11},\quad \bm{\Gamma}=(\mathcal{M}_p^{-1}\bar{\mathbf{u}}_0^>,-\bar{\mathbf{u}}_0^<).\label{eq:amplitude}
\end{align}
The net transmission probability for the two combined barriers can be written in a general Fabry-P\'erot form \cite{datta1997electronic,PhysRevLett.101.156804}
\begin{align}
T=\frac{t_Lt_R}{|1-\sqrt{r_Lr_R}e^{i\Delta\phi}|^2},\label{Eq:tt}
\end{align}
where $r_{L(R)}=1-t_{L(R)}$ is the reflection probability for the left (right) barrier and $\Delta\phi$ is the total phase shift acquired through the tunneling process.

The transverse and the longitudinal currents can be calculated within the Landauer formalism \cite{PhysRevB.24.2978,datta1997electronic}, which are given by
\begin{align}
I_{yx}^\tau=&\frac{e}{h}\sum_{k_y}\int_{-\infty}^{\infty}dE\ \Big(\frac{v_{y}^>}{v_{x}^>}-\frac{v_{y}^<}{v_{x}^>}R\Big)[f(E-eV)-f(E)],\\
I_{xx}^\tau=&\frac{e}{h}\sum_{k_y}\int_{-\infty}^\infty dE\ \Big(1+\frac{v_{x}^<}{v_{x}^>}R\Big)[f(E-eV)-f(E)],
\end{align}
respectively. Here $\tau=\pm1$ is the valley index, $R=1-T$ is the reflection probability, $V$ is the longitudinal voltage drop along the junction, $f(E)=1/(\mathrm{exp}(E/k_BT')+1)$ is the Fermi-Dirac distribution function with $k_B$ and $T'$ being the Boltzmann constant and temperature, respectively, and $v^\varrho_{\eta}=\langle\mathbf{u}_0^\varrho|\hat{v}_\eta|\mathbf{u}_0^\varrho\rangle$ is the group velocity along the $\eta$ axis ($\eta=x,y$) for the propagating state $\mathbf{u}_0^\varrho$ ($\varrho=>,<$) with the velocity operator $\hat{v}_\eta=\partial\mathcal{H}/\partial \hbar k_\eta$. The zero-temperature transverse and the longitudinal conductance for the $\tau$ valley can be obtained by $\partial I_{\eta x}^\tau/\partial(eV)$ at $T'=\SI{0}{\K}$, which are given by
\begin{align}
\sigma_{yx}^\tau&=\sigma_0\int_{-\pi/2}^{\pi/2}d\theta_0\ T_{\tau}(\theta_0)\sin\theta_0,\label{eq:cttx}\\
\sigma_{xx}^\tau&=\sigma_0\int_{-\pi/2}^{\pi/2}d\theta_0\ T_{\tau}(\theta_0)\cos\theta_0,\label{eq:cll}
\end{align}
respectively. Here $\sigma_0=(2e^2/h)N_0(E)$ is the normalized conductance, where $N_0(E)=(E-V_p)W/\pi\hbar v_F$ is the number of transverse modes with $W$ being the junction width. Consequently, the net longitudinal conductance is given by
\begin{align}
\sigma_{xx}=\sigma_{xx}^K+\sigma_{xx}^{K^\prime}.\label{eq:ccx}
\end{align} 
The net transverse charge conductance and transverse valley conductance are given by 
\begin{align}
\sigma_{yx}&=\sigma_{yx}^K+\sigma_{yx}^{K^\prime},\quad \sigma^{V}_{yx}=\sigma_{yx}^{K}-\sigma_{yx}^{K^\prime},\label{eq:ccy}
\end{align} 
respectively. The charge Hall angle and the valley Hall angle can be obtained by
\begin{align}
\tan\Theta=\frac{\sigma_{yx}}{\sigma_{xx}},\quad \tan\Theta_{V}=\frac{\sigma_{yx}^V}{\sigma_{xx}},\label{eq:vha}
\end{align}
respectively.

\begin{figure}[tp]
\begin{center}
\includegraphics[clip = true, width =\columnwidth]{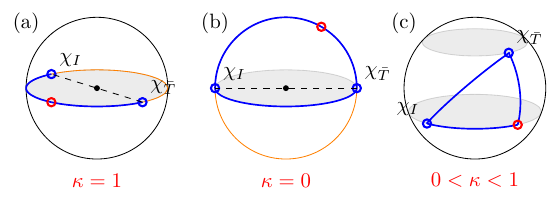}
\end{center}
\caption{The geodesic polygon on the Bloch sphere. Two blue vertices denote the incident state $\chi_I$ and the transmitted state $\chi_{\bar{T}}$. The red vertex denotes the reflected state $\chi_R$. (a) Graphene lattice ($\kappa=1$). (b) Dice lattice ($\kappa=0$). (c) $\alpha-\mathcal{T}_3$ lattice ($0<\kappa<1$).}
\label{fig:sphere}
\end{figure}

\section{Results}\label{results}
\subsection{Valley-dependent phase shift} 
\label{sub:Valley-dependent geometric phase coherence}
The phase of the backreflection amplitude acquired at the interface of the $p$ barrier ($p=L,R$) can be obtained by the current conservation condition
\begin{align}
\mathcal{M}_p'(\chi_{I}+\sqrt{r_p}e^{i\phi_p}\chi_{R})=\sqrt{t_p}e^{i\gamma_p}\chi_{T},\label{eq:cc}
\end{align}
where the transfer matrix is $\mathcal{M}_p'=\mathcal{M}_p$ for $p=R$ and $\mathcal{M}_p'=\mathcal{M}_p^{-1}$ for $p=L$. $\chi_{I}$, $\chi_{R}$ and $\chi_{T}$ are the incident, reflected and transmitted pseudospinors, respectively. $\phi_p$ ($\gamma_p$) is the phase of the reflection (transmission) amplitude. The reflection amplitude can be directly obtained by solving Eq.\ (\ref{eq:cc}), resulting in $\sqrt{r_p}e^{i\phi_p}=-\chi_{\bar{T}}^\dagger\mathcal{M}'_p\chi_{I}(\chi_{\bar{T}}^\dagger\mathcal{M}'_p\chi_{R})^{-1}$, where $\chi_{\bar{T}}$ is the pseudospinor orthogonal to $\chi_{T}$. With the help of the geodesic rule \cite{PhysRevB.89.155412,doi:10.1080/09500348714551321} of $\arg(\chi_a^\dagger\chi_b)=i\int_\mathcal{C}d\mathbf{s}\cdot\chi^\dagger_\mathbf{s}\bm\nabla\chi_\mathbf{s}$ (the integration path $\mathcal{C}$ is along the geometric line from $\chi_b$ to $\chi_a$ on the Bloch sphere), the phase of the backreflection amplitude is given by
\begin{align}
    \phi_p=\pi-\arg(\chi_{I}^\dagger\chi_{R})+\phi_{M,p}+\phi_{G,p},\label{eq:phi}
\end{align}
where $\phi_{M,p}$ is the Wentzel–Kramers–Brillouin (WKB) phase \cite{PhysRevA.40.6814,PhysRev.41.713} accumulated in the barrier region and $\phi_{G,p}=-\Omega_{I\bar{T}R}/2$ is the geometric phase acquired through the tunneling process of barrier $p$ with $\Omega_{I\bar{T}R}$ being the solid angle covered by the geodesic triangle connecting the states $\chi_I$, $\chi_{\bar T}$ and $\chi_R$ on the Bloch sphere \cite{PhysRevB.89.155412,PhysRevB.87.165420}. The second term in Eq.\ (\ref{eq:phi}) is gauge-dependent. In the magnetic-field free Fabry-P\'erot model, $\arg(\chi_{I}^\dagger\chi_{R})$ at $L$ and $R$ interfaces are equal in magnitude but opposite in sign \cite{PhysRevLett.101.156804}. Consequently, the total phase shift is given by
\begin{align}
    \Delta\phi=2k_0l+\sum_p\phi_p=2k_0l+\sum_p\phi_{M,p}+\sum_p\phi_{G,p},\label{eq:pph}
\end{align}
where the first term $2k_0l$ is the WKB phase acquired at the central spacer region and $\sum_p\phi_p$ is the total phase acquired at the left and right barrier regions.  Eq.\ (\ref{eq:pph}) can be divided into two parts, namely
\begin{align}
\Delta\phi=\phi_{\mathrm{WKB}}+\phi_G,\label{eq:totp}
\end{align} 
where $\phi_{\mathrm{WKB}}$ is the total WKB phase and $\phi_G$ is the total geometric phase.

\subsubsection{WKB phase}
The total WKB phase in Eq.\ (\ref{eq:pph}) is given by 
\begin{align}
    \phi_{\mathrm{WKB}}=2k_0l+\sum_p\phi_{M,p},\label{eq:wkb}
\end{align}
where the first term ($2k_0l$) is the WKB phase accumulated in the central spacer region and the second term is the total WKB phase accumulated in the left and right barrier regions. One finds 
\begin{align}
\tan\phi_{M,p}=-\frac{\cos^2\theta_0+\cos^2\theta_p+\kappa^2(\sin\theta_0-\sin\theta_p)^2}{2\cot k_pd\cos\theta_0\cos\theta_p},\label{eq:mp}
\end{align}
where $\kappa=(1-\alpha^2)/(1+\alpha^2)$ is the magnitude of the Berry phase (in units of $\pi$). We note that $k_0=E\cos\theta_0/\hbar v_F$, $k_p=(E-V_p)\cos\theta_p/\hbar v_F$, and $\sin\theta_p=E\sin\theta_0/(E-V_p)$. As a result, $k_0$ and $k_p$ are an even function of $\theta_0$ whereas $\theta_p$ is an odd function of $\theta_0$. 

With the help of Eqs.\ (\ref{eq:wkb}-\ref{eq:mp}), it is found that the total WKB phase is valley-independent and symmetric with the substitution $\theta_0\rightarrow-\theta_0$.

\subsubsection{Geometric phase}
The total geometric phase in Eq.\ (\ref{eq:pph}) is given by $\phi_G=\sum_p\phi_{G,p}$, where $\phi_{G,p}$ is the geometric phase acquired through the tunneling process in the $p$ ($p=L,R$) barrier. One finds 
\begin{align}
    \tan\phi_{G,p}=(-1)^{\ell_p}\tau\frac{1-\kappa^2}{2\kappa}\frac{\sin\theta_p}{\cos\theta_0},\label{eq:gp0}
\end{align}
where $\tau=\pm1$ labels two different valleys and $\ell_p=+1(-1)$ for $p=L(R)$. The total geometric phase $\phi_G=\sum_p\phi_{G,p}$ is given by
\begin{align}
\tan\phi_G=\tau\frac{2\kappa(\kappa^2-1)(\sin\theta_L-\sin\theta_R)}{4\kappa^2\cos\theta_0+(1-\kappa^2)^2\sec\theta_0\sin\theta_L\sin\theta_R}.\label{eq:gp}
\end{align}
Distinct from the WKB phase, the total geometric phase $\phi_G$ is valley-dependent and asymmetric with respect to the incident angle $\theta_0$. Specifically, $\phi_G$ reverses its sign with the substitution $\theta_0\rightarrow-\theta_0$ or $\tau\rightarrow-\tau$. 

\begin{figure}[tp]
\begin{center}
\includegraphics[clip = true, width =\columnwidth]{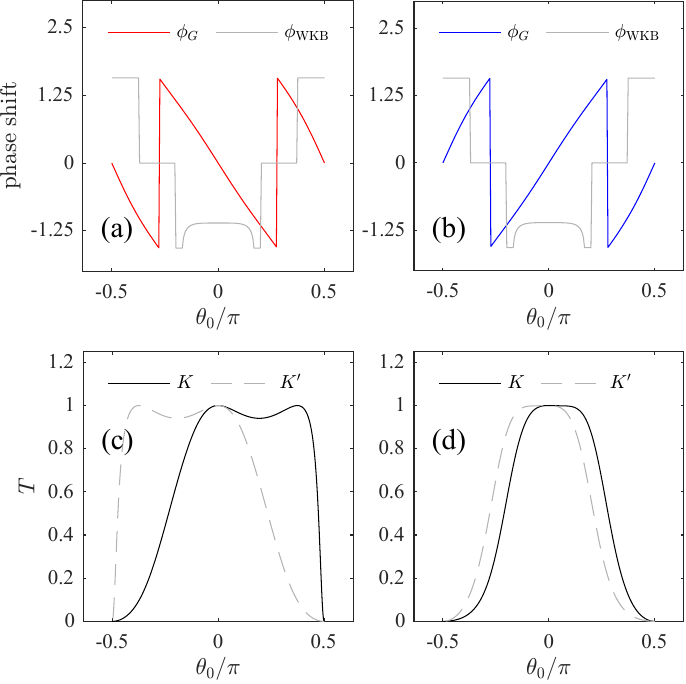}
\end{center}
\caption{(a) The phase shift versus the incident angle for $K$ valley. The geometric phase and the WKB phase are denoted by the red and gray lines, respectively. (b) The phase shift versus the incident angle for $K^\prime$ valley. The geometric phase and the WKB phase are denoted by the blue and gray lines, respectively. (c) Transmission probability versus the incident angle for $K$ valley (solid) and $K^\prime$ valley (dashed) at $\alpha=0.4$. (d) Transmission probability versus the incident angle for $K$ valley (solid) and $K^\prime$ valley (dashed) at $\alpha=0.7$. The heights of two barriers are set to $V_L=\SI{0.2}{\eV}$ and $V_R=\SI{1.2}{\eV}$, respectively. The width of the barrier is $d=\SI{25}{\nm}$ and the width between two barriers is $l=\SI{50}{\nm}$.}
\label{fig:phase}
\end{figure}

From Eq.\ (\ref{eq:gp}), it is found that the total geometric phase is always absent for the integer Berry phase $\kappa=0$ and $\kappa=1$, corresponding to the dice lattice and graphene lattice, respectively. For the non-integer Berry phase, the nonzero $\phi_G$ appears for the asymmetric potential barriers ($V_L\neq V_R$) where $\theta_L\neq\theta_R$. 

The behavior of $\phi_G$ mentioned above can be understood geometrically as follows. The scattering states in Eq.\ (\ref{eq:cc}) are two-component spinors and can be mapped onto the Bloch sphere with spin-$1/2$ Pauli matrices. Taking the interface of the right barrier as an example. The vectors on the Bloch sphere corresponding to $\chi_I$ ($\chi_R$) and $\chi_{\bar T}$ are given by
\begin{align}
&\mathbf{v}_{I(R)}=\rho_0^{-1}\big(\pm4\cos\theta_0,4\kappa\tau\sin\theta_0,2(\kappa^2-1)\sin^2\theta_0\big),\label{eq:vv1}\\
&\mathbf{v}_{\bar T}=\rho_R^{-1}\big(-4\cos\theta_R,-4\kappa\tau\sin\theta_R,2(1-\kappa^2)\sin^2\theta_R\big),\label{eq:vv2}
\end{align}
respectively, where $\rho_{0(R)}=4-2(1-\kappa^2)\sin^2\theta_{0(R)}$ is the normalized factor. For $\kappa=1$ (graphene lattice), $\chi_I$, $\chi_R$ and $\chi_{\bar T}$ lie on the equator of the Bloch sphere, as shown in Fig.\ \ref{fig:sphere}(a). The area of the geodesic triangle connecting $\chi_I$, $\chi_R$ and $\chi_{\bar T}$ is always zero, leading to the zero solid angle as well as the geometric phase. For $\kappa=0$ (dice lattice), $\chi_I$, $\chi_R$ and $\chi_{\bar T}$ lie on the $v_y=0$ plane of the Bloch sphere. For the electrons from $K$ valley ($\tau=+1$), the geodesic triangle connecting $\chi_I$, $\chi_R$ and $\chi_{\bar T}$ covers a quarter of the Bloch sphere with the solid angle being $4\pi/4=\pi$, as shown in Fig.\ \ref{fig:sphere}(b). Consequently, the geometric phase is $\phi_{G,R}=-\pi/2$, which is in agreement with Eq.\ (\ref{eq:gp0}) by substituting $\kappa\rightarrow0$. For $\kappa\neq0,1$, the vectors $\mathbf{v}_{I(R)}$ and $\mathbf{v}_{\bar T}$ lie on the planes parallel to the equatorial plane with the distance to the equatorial plane being $1-4/\rho_{0}$ and $1-4/\rho_{R}$, respectively. Consequently, the nonzero solid angle covered by the geodesic polygon connecting $\chi_{I}$, $\chi_{R}$ and $\chi_{\bar T}$ appears, as shown in Fig.\ \ref{fig:sphere}(c), which is dependent on $\theta_0$ and $\theta_R$ and can be modified by the junction control. Swapping the valley index changes the sign of the second components of the vectors in Eqs.\ (\ref{eq:vv1}-\ref{eq:vv2}), leading to the sign reversal of the geometric phase in Eq.\ (\ref{eq:gp0}). The geometric phase acquired at the left interface of the barrier can be similarly obtained by the substitution $\theta_0\rightarrow-\theta_0$ and $\theta_R\rightarrow-\theta_L$. For the symmetric potential ($\theta_L=\theta_R$), the acquired geometric phases at the left and right barrier interfaces are equal in magnitude but opposite in sign, which result in the zero total geometric phase in Eq.\ (\ref{eq:gp}).

The phase shift as a function of the incident angle is shown in Figs.\ (\ref{fig:phase})(a) and (\ref{fig:phase})(b) for $K$ and $K^\prime$ valleys, respectively. It is shown that the WKB phase is valley-independent and is symmetric with respect to the incident angle. However, the geometric phase is opposite for different valleys and is asymmetric with respect to the incident angle. This valley-dependent geometric phase shift plays a key role in the valley-contrasting coherent transport in our model, which is explained in detail in Sec.\ \ref{sub:Tunneling valley Hall effect}.

\subsection{Tunneling valley Hall effect} 
\label{sub:Tunneling valley Hall effect}
With the help of Eqs.\ (\ref{eq:totp}-\ref{eq:gp}), the valley dependence of the total phase shift can be summarized in the following compact form
\begin{align}
\Delta\phi=\phi_{\mathrm{WKB}}+\tau\times\mathrm{sgn}(\theta_0)|\phi_G|,
\end{align}
where $\tau=\pm1$ labels the $K$ and $K^\prime$ valleys, respectively. Both $\phi_\mathrm{WKB}$ and $|\phi_G|$ are valley-independent and even functions of the incident angle ($\theta_0$). The transmission probability of the phase coherent transport can be obtained by Eq.\ (\ref{Eq:tt}), which is given by
\begin{gather}
T=\frac{t_Lt_R}{1-2\sqrt{r_Lr_R}\cos\big(\phi_{\mathrm{WKB}}+\tau\mathrm{sgn}(\theta_0)|\phi_G|\big)+r_Lr_R}.\label{eq:tot}
\end{gather}
The single barrier tunneling transmission probability in Eq.\ (\ref{eq:tot}) is given by
\begin{align}
t_p=\left|\frac{2\cos\theta_0\sec\theta_p}{w\sin k_pd+2i\cos\theta_0\sec\theta_p\cos k_pd}\right|^2,
\end{align}
where $p=L,R$ for the left and right barriers, respectively, $w=\sec^2\theta_p(\cos^2\theta_0+\kappa^2(\sin\theta_0-\sin\theta_p)^2)+1$, and the reflection probability is given by $r_p=1-t_p$ due to the conservation of the probability current. For the single rectangular barrier tunneling, both $t_p$ and $r_p$ are valley-independent and are even functions of the incident angle $\theta_0$ \cite{PhysRevB.95.235432,PhysRevB.105.165402,PhysRevB.103.165429}. As a result, the valley dependence and the asymmetry of $\theta_0$ are attributed to the coherent geometric phase.  

\begin{figure}[tp]
\begin{center}
\includegraphics[clip = true, width =\columnwidth]{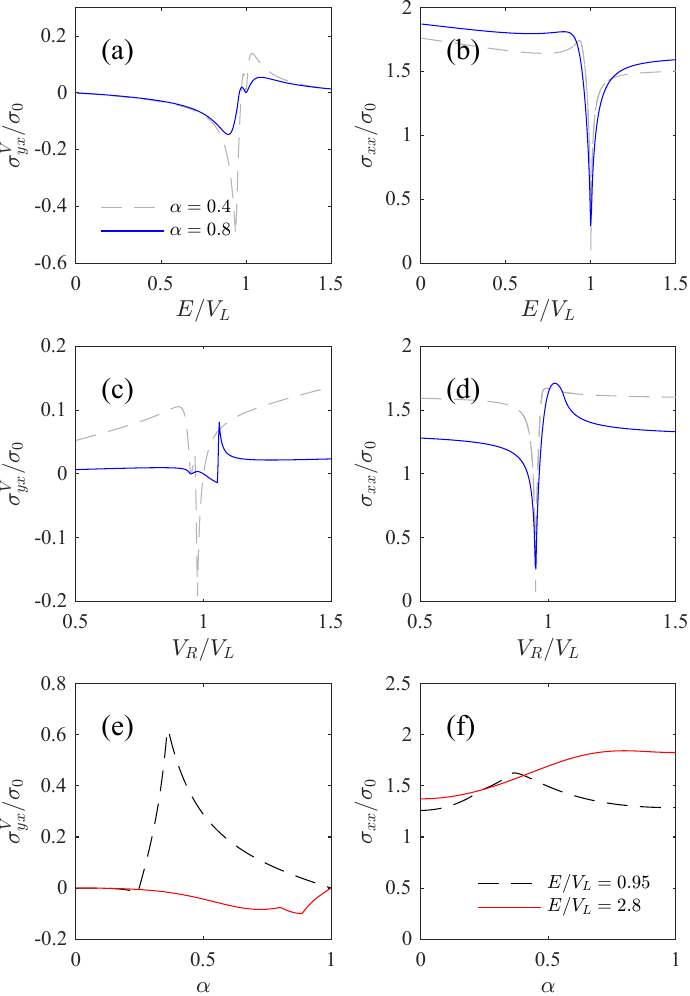}
\end{center}
\caption{(a) Transverse valley conductance as a function of the incident energy at $\alpha=0.4$ (gray dashed) and $\alpha=0.8$ (blue solid). The heights of two barriers are $V_L=\SI{0.2}{\eV}$ and $V_R=\SI{1.2}{\eV}$. The width of the barrier is $d=\SI{25}{\nm}$ and the width between two barriers is $l=\SI{50}{\nm}$. (b) Longitudinal conductance as a function of the incident energy. The parameters are the same as (a). (c) Transverse valley conductance as a function of the height of the right barrier $V_R$. The height of the left barrier is set to $V_L=\SI{0.2}{\eV}$. The incident energy is set to $E=0.95V_L$. The other parameters are the same as (a). (d) Longitudinal conductance as a function of the height of the right barrier $V_R$. The parameters are the same as (c). (e) Transverse valley conductance as a function of the lattice parameter $\alpha$ at $E=0.95V_L$ (black dashed) and $E=2.8V_L$ (red solid). The other parameters are the same as (a). (f) Longitudinal conductance as a function of the lattice parameter $\alpha$, the parameters are the same as (e).}
\label{fig:conductance}
\end{figure}

The transmission probability as a function of the incident angle is shown in Fig.\ \ref{fig:phase}(c), where $\alpha=0.4$, the heights of the barriers are set to $V_L=\SI{0.2}{\eV}$ and $V_R=\SI{1.2}{\eV}$. It is shown that the electrons in $K$ valley have a large transmission probability for the incident angle in the range of $0^\circ$ to $90^\circ$ (black solid line), whereas the transmissions in $K^\prime$ valley are similarly asymmetric but skewed into the opposite direction (black dashed line). This valley-contrasting skew tunneling leads to the carries in different valleys turning into different transverse directions, which is responsible for the nonzero transverse valley currents. This similar valley-contrasting skew tunneling also occurs for $\alpha=0.7$, as shown in Fig.\ \ref{fig:phase}(d). In fact, for $\phi_G\neq0$, the valley-dependent skew tunneling always occurs, \textit{i.e.}, $T_\tau(\theta_0)\neq T_\tau(-\theta_0)$; see Eq.\ (\ref{eq:tot}). Due to the time-reversal symmetry, the total transmission probability is always symmetric with respect to the incident angle, \textit{i.e.}, $T_K(\theta_0)+T_{K^\prime}(\theta_0)=T_K(-\theta_0)+T_{K^\prime}(-\theta_0)$, which results in the symmetric patterns between $K$ and $K^\prime$ valleys [$T_K(\theta_0)=T_{K^\prime}(-\theta_0)$] in Figs.\ \ref{fig:phase}(c-d) and is responsible for zero transverse charge currents.

The longitudinal conductance, transverse charge conductance and transverse valley conductance can be obtained by Eqs.\ (\ref{eq:ccx}-\ref{eq:ccy}). Due to the time-reversal symmetry, the symmetric relation of the transmission probability between two valleys is always valid, \textit{i.e.}, $T_K(\theta_0)=T_{K^\prime}(-\theta_0)$, which leads to the transverse charge conductance of $K$ and $K^\prime$ valleys satisfies $\sigma_{yx}^{K}=-\sigma_{yx}^{K^\prime}$; see Eq.~(\ref{eq:cttx}). Consequently, the net transverse charge conductance $\sigma_{yx}=\sigma_{yx}^{K}+\sigma_{yx}^{K^\prime}$ is always zero. 

The transverse valley conductance ($\sigma^{V}_{yx}$) and the longitudinal conductance ($\sigma_{xx}$) versus the incident energy are shown in Figs.\ \ref{fig:conductance}(a) and \ref{fig:conductance}(b), respectively, where the height of the left barrier is set to $V_L=\SI{0.2}{\eV}$. Both $\sigma^{V}_{yx}$ and $\sigma_{xx}$ are absent at $E=V_L$ due to the zero density of states at the Dirac point. For $E<V_L$, the transmitted electrons in the left barrier lie in the valence band, where the right propagating states have negative longitudinal wave vectors. However, for $E>V_L$, the transmitted carriers are the electrons in the conduction band with positive longitudinal wave vectors. This sign change of the longitudinal wave vectors lead to the reversal of $\sigma^{V}_{yx}$, indicating the reversal of the transverse valley currents, as shown in Fig.\ \ref{fig:conductance}(a). $\sigma^{V}_{yx}$ and $\sigma_{xx}$ versus the height of the right barrier are shown in Figs.\ \ref{fig:conductance}(c) and \ref{fig:conductance}(d), respectively, where the incident energy is set to $E=0.95V_L$. It is shown that the transverse valley conductance disappears at $V_R/V_L=1$, due to the absence of the total geometric phase. At the Dirac point of the right barrier region, where $V_R/V_L=E/V_L=0.95$, both the transverse valley conductance and the longitudinal conductance are absent. $\sigma^{V}_{yx}$ and $\sigma_{xx}$ versus the lattice parameter $\alpha$ are shown in Figs.\ \ref{fig:conductance}(e) and \ref{fig:conductance}(f), respectively. The transverse valley conductance is always absent at $\alpha=0$ (graphene lattice) and $\alpha=1$ (dice lattice), where the total geometric phase is zero; see Fig.\ \ref{fig:conductance}(e). However, the longitudinal conductance is finite for all $\alpha$; see Fig.\ \ref{fig:conductance}(f).

\begin{figure}[tp]
\begin{center}
\includegraphics[clip = true, width =\columnwidth]{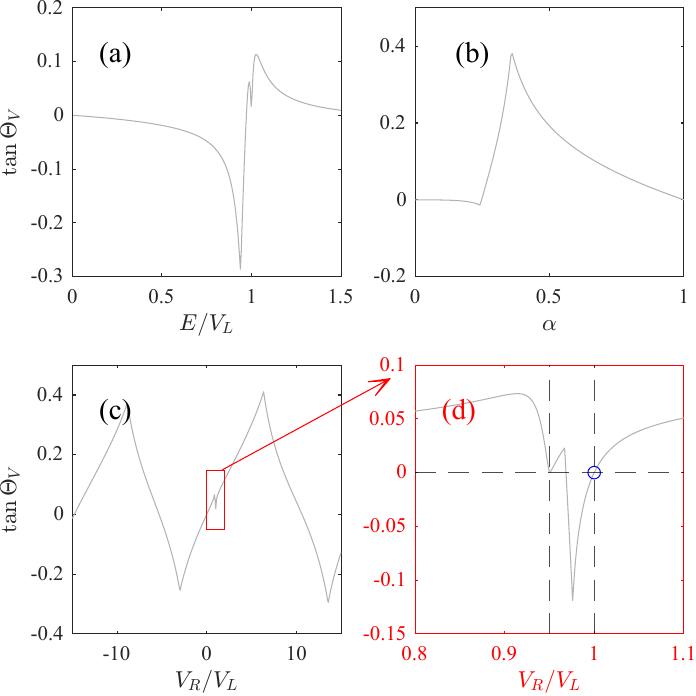}
\end{center}
\caption{(a) The tangent of the valley Hall angle $\tan\Theta_{\mathrm{V}}$ as a function of the incident energy $E$ at $\alpha=0.4$ and $V_R/V_L=-3.5$. (b) The tangent of the valley Hall angle $\tan\Theta_{\mathrm{V}}$ versus $\alpha$ at $V_R/V_L=-3.5$. (c) The tangent of the valley Hall angle $\tan\Theta_{\mathrm{V}}$ versus $V_R/V_L$ at $\alpha=0.4$. (d) Zoom in on the red box of panel (c). The heights of the two barriers are $V_L=\SI{0.2}{\eV}$ and $V_R=\SI{1.2}{\eV}$. The width of the barrier is $d=\SI{25}{\nm}$ and the width between two barriers is $l=\SI{50}{\nm}$.}
\label{fig:vv}
\end{figure}

The efficiency of the charge-valley conversion is characterized by the valley Hall angle $\Theta_{V}$, which can be obtained by Eq.\ (\ref{eq:vha}). The tangent of the valley Hall angle ($\tan\Theta_{V}$) as a function of the incident energy ($E$) is shown in Fig.\ \ref{fig:vv}(a), where $\alpha=0.4$, $V_L=\SI{0.2}{\eV}$ and $V_R=-3.5V_L$. The absolute value of the valley Hall angle increases with the increasing of the incident energy, and changes its sign at the Dirac point of the left barrier region \textit{i.e.}, $E= V_L$, indicating the reversal of the transverse valley Hall currents. The tangent of the valley Hall angle as a function of the parameter $\alpha$ is shown in Fig.\ \ref{fig:vv}(b).  It is shown that $\tan \Theta_{V}$ approaches its maximum value at $\alpha\simeq0.35$, and vanishes at $\alpha=0$ and $\alpha=1$, indicating the absence of the transverse valley Hall currents in the graphene lattice and the dice lattice, respectively. The transverse valley Hall currents can be electrically controlled by the gate voltages applied across the two barrier regions. The valley Hall angle as a function of $V_R/V_L$ is shown in Figs.\ \ref{fig:vv}(c) and \ref{fig:vv}(d), where the potential height of the left barrier region is set to $V_L=\SI{0.2}{\eV}$ and the incident energy is set to $E=0.95V_L$ . Due to the coherence of the geometric phase, the valley Hall angle varies periodically with increasing $V_R$, as shown in Fig.\ \ref{fig:vv}(c). The more detailed information around the regime of $V_R/V_L=1$ [red box in Fig.\ \ref{fig:vv}(c)] is shown in Fig.\ \ref{fig:vv}(d), where $0.8V_L<V_R<1.1V_L$. It is shown that the transverse valley Hall current is absent at $V_R=V_L$, which is attributed to the absence of the total geometric phase. For $V_R=E=0.95V_L$, the incident energy approaches the Dirac point of the right barrier region, due to the absence of the density of states, the transverse valley Hall current also vanishes, leading to $\tan\Theta_{V}=0$.

\section{Conclusions}\label{conclusions}
To conclude, we theoretically investigate the coherent transport through two combined electric barriers in $\alpha-\mathcal{T}_3$ lattices. It is shown that the backreflected electron may acquire a valley-dependent geometric phase in the single tunneling process through the rectangular barrier of the $\alpha-\mathcal{T}_3$ lattice. This acquired geometric phase is determined by the height of the barrier, which can be electrically controlled. The coherence of this geometric phase in this double barrier structure leads to the asymmetry of the valley-dependent transmission probability, which is responsible for the transverse valley currents with zero net charge. We further demonstrate that this charge-neutral transverse valley Hall current can be electrically controlled by the gate voltages applied across the left ($V_L$) and right ($V_R$) barrier regions, and disappears at $V_L=V_R$.

\end{document}